\documentclass[a4paper,10pt,twoside]{article}

\usepackage[latin1]{inputenc} 
\usepackage[T1]{fontenc} 

\usepackage{amsmath}
\usepackage{amssymb}
\usepackage{graphicx}
\usepackage{color}
\usepackage{epic}
\usepackage{a4wide}
\usepackage{enumerate}

\title{A reduced model for shock and detonation waves. II. The reactive case.}
\author{J.B. Maillet$^1$, L. Soulard$^1$, and G. Stoltz$^{1,2}$
  \bigskip\\
  $^1$ CEA/DAM Ile-de-France,
  BP 12, 91680 Bruy\`eres-le-Ch\^atel, France. \\
  $^2$ CERMICS, Ecole Nationale des
  Ponts et Chauss\'ees (ParisTech), \\ 6 \& 8 Av. Pascal,
  77455 Champs-sur-Marne, France. \\
\\
{\tt \{jean-bernard.maillet,laurent.soulard\}@cea.fr, stoltz@cermics.enpc.fr}
}

\begin{document}

\maketitle

\abstract{We present a mesoscopic model for reactive shock waves, which
extends the model proposed in~\cite{Stoltz06}. A complex molecule
(or a group of molecules) is replaced by a single mesoparticle,
evolving according to some Dissipative Particle Dynamics. Chemical
reactions can be handled in a mean way by considering an
additional variable per particle describing the progress of the reaction.
The evolution of the progress variable is governed by the kinetics of a
reversible exothermic reaction. Numerical results give profiles in
qualitative agreement with all-atom studies.}

\section{Introduction}

Whereas multimillion atoms simulations are nowadays common in
molecular dynamics studies with simple potentials, the time and
space scales numerically tractable are still far from being
macroscopic. The situation is even worse for nonequilibrium
processes such as shock and detonation waves. Indeed, the
simulation of detonation requires the description of a thin shock
front, moving at high velocity, usually using a complicated
empirical potential able to treat chemical events happening
(dissociation, recombination - see~\cite{FD} for a fundamental
reference). To this end, toy molecular models were proposed at the
early stages of molecular simulations of detonation (see {\it
e.g.}~\cite{POLS85}), until the first all-atom studies in the
nineties~\cite{BREW93,BREW96}, allowed by the development of bond
order potentials. Nevertheless, despite recent advances in the
development of reactive force fields~\cite{Nielson05}, the
simulation of the decomposition of real materials still remains
confined to small systems sizes and short time scales (see for
example~\cite{SvDCDG03} for a state of the art study). A direct
simulation of the macroscopic reaction zone of a real material is
well beyond current computer capabilities, highlighting the
importance of the development of reduced model for detonation.

Following the pioneering works of Holian and
Strachan~\cite{Holian03,SH04}, a reduced model for (inert) shock
waves was proposed in~\cite{Stoltz06}, where complex molecules are
replaced by mesoparticles. These mesoparticles are described by
their positions, momenta, and have an additional degree of
freedom: their internal energies. This model is strongly inspired
by Dissipative Particle Dynamics (DPD)~\cite{HK92} with conserved
energy~\cite{AM97,Espanol97}. In reduced models for
shock waves~\cite{SH04,Stoltz06}, one
mesoparticle stands for one complex molecule. Reduced models are
interesting in practice to simulate large systems, and as
an intermediate step in a truly multiscale approach, where some parts
of the system could be treated with all-atom models, while the
remaining parts would be treated with hydrodynamic models,
implemented with particle discretizations such as Smoothed
Particle Hydrodynamics~\cite{Lucy77,Monaghan92}. A first step to
such a general formalism in the equilibrium case is proposed
in~\cite{ER03}.

In the reactive case, exothermic chemical reactions are triggered
when the shock passes, and the energy liberated sustains the
shock. To model detonation at the mesoscopic level, we introduce a
new variable per mesoparticle, namely a progress variable, which
characterizes the progress of the chemical decomposition. The
dynamics can then be split into three elementary physical
processes: 
\begin{enumerate}[\quad (i)]
\item the translational dynamics of the particles, given
by the dynamics of inert materials~\cite{Stoltz06}; 
\item the
evolution of chemical reactions through some kinetics on the progress variable; 
\item the exothermicity of the reaction: energy transfers between chemical
and mechanical plus internal energies have to be precised.
\end{enumerate}

The paper is organized as follows. After summarizing the dynamics
for inert materials, we turn to the evolution of the progress variable
and the treatment of the exothermicity in the reactive case. A
numerical implementation relying on some splitting based on the
decomposition of the dynamics into elementary physical processes
is also proposed. Numerical results eventually confirm the correct
behavior of the model. Let us emphasize that we aim here at proposing
a dynamics in qualitative agreement with all-atom detonation
simulations, giving correct orders of magnitude for the speed of the
detonation front and the width of the reaction zone. This dynamics is of course
parametrized by a certain collection of parameters describing the initial
and the final material, as well as the chemical kinetics. We however
tried to limit their total number to keep only the
essential ones, and give some tracks to estimate these parameters
from small all-atom numerical simulations or physical experiments.


\section{A reduced model for inert shock waves}
\label{sec:inert}

We briefly recall here the inert model presented
in~\cite{Stoltz06}. We consider a system of $N$ mesoscopic
particles with positions $(q_1,\dots,q_N)$, momenta
$(p_1,\dots,p_N)$, masses $(m_1,\dots,m_N)$, and internal energies
$(\epsilon_1,\dots,\epsilon_N)$. The internal temperature $T_i$ is
defined as $T_i^{-1} = \frac{ds(\epsilon_i)}{d\epsilon_i}$.
The function $s \equiv s(\epsilon)$ is the microscopic
equation of state of the system,
and relates the microscopic entropy~$s$ (arising from the existence of
degrees of freedom not explicitely represented)
and the internal energy~$\epsilon$.
As a first approximation,
$s(\epsilon) = C_v \ln \epsilon$. Denoting by $\bar{T}$ the
reference temperature, $\beta = 1/(k_{\rm B} \bar{T})$, and $V$ the
interaction potential, the system evolves in the inert case as
\begin{eqnarray}
\label{DPDsimplifie}
dq_i & = & \frac{p_i}{m_i} \, dt, \nonumber \\
dp_i & = & \sum_{j, \, j \not = i} -\nabla_{q_i} V(r_{ij}) \, dt
- \gamma_{ij} \chi^2(r_{ij}) v_{ij} \, dt \nonumber \\
& & \quad + \sigma \chi(r_{ij}) dW_{ij}, \\
d\epsilon_i & =& \frac12 \sum_{j, \, j \not = i} \chi^2(r_{ij}) \left (
\gamma_{ij} v_{ij}^2 - \frac{d\sigma^2}{2}
\left(\frac{1}{m_i}+\frac{1}{m_j}\right) \right ) dt
- \sigma \, \chi(r_{ij}) v_{ij} \cdot dW_{ij}, \nonumber
\end{eqnarray}
where $r_{ij} = |q_i-q_j|$, $v_{ij}=\frac{p_i}{m_i}-\frac{p_j}{m_j}$,
$\chi$ is a weighting function (with support
in $[0,r_{\rm c}]$ for some cut-off radius $r_{\rm c}$), and the
processes $W_{ij}$ are independent $d$-dimensionnal standard
Wiener processes such that $W_{ij}=-W_{ji}$. The
fluctuation-dissipation relation relating the frictions
$\gamma_{ij}$ and the magnitude $\sigma$ of the random forcing is
\begin{equation}
\label{eq:FDR}
\gamma_{ij} = \sigma^2 \beta_{ij}/2, \quad \beta_{ij} =
\frac{1}{k_{\rm B}T_{ij}} = \frac{1}{2 k_{\rm B}} \left (
\frac{1}{T_i} + \frac{1}{T_j} \right ).
\end{equation}
Notice that we implicitely defined the temperature~$T_{ij}$ as the
average internal temperature: $T_{ij} = (T_i+T_j)/2$.
It can be shown that the measure
\[
d\rho(q,p,\epsilon) = Z^{-1} {\rm e}^{-\beta H(q,p)} \exp\left(\sum_i
\frac{s(\epsilon_i)}{k_{\rm B}}-\beta\epsilon_i\right) \delta_{E=E_0}
\, \delta_{P=P_0} \, \, dq \, dp \, d\epsilon
\]
is an invariant measure of the dynamics.
In practice, we set $\sigma^2 = 2\gamma/\beta$ and therefore
$\gamma_{ij}/\gamma = \bar{T}/T_{ij}$. The parameters used in the
inert dynamics can be estimated as in~\cite{Stoltz06}.


\section{Evolution of the progress variable}

In the reactive case, chemical reactions are triggered when the
shock passes. To model the progress of the reaction, an additional
degree of freedom, a progress variable~$\lambda_i$, is attached to
each particle. For the model reaction
\begin{equation}
\label{eq:reaction_chimique}
2 \text{AB} \rightleftarrows \text{A}_2 + \text{B}_2,
\end{equation}
the state $\lambda = 0$ corresponds to a molecule AB, whereas the
state $\lambda = 1$ corresponds to A$_2$ + B$_2$. Representing the
progress of the chemical reaction by a real-value parameter
seems questionable when a mesoparticle stands for a single
molecule. The progress variable should therefore rather be seen as some
dissociation probability, or progress along some free energy
profile.

Since the model reaction~(\ref{eq:reaction_chimique}) involves two
species on each side, we postulate for example the reversible
evolution:
\begin{eqnarray}
\label{eq:chemical_reaction}
\frac{d\lambda_i}{dt} & = & \sum_{j \not = i} \omega_r(r_{ij}) \left [
  K_1(T_{ij}) ( 1 - \lambda_i)(1-\lambda_j) + K_2(T_{ij}) \lambda_i \lambda_j \right ]
\end{eqnarray}
when~$0 \leq \lambda_i \leq 1$, and $\frac{d\lambda_i}{dt} = 0$
otherwise (in order to ensure that the progress variables remains
in~$[0,1]$). The function $\omega_r$
in~\eqref{eq:chemical_reaction} is a weight function (with support
in~$[0,r_{\rm reac}]$), and the mean temperature $T_{ij}$ is
defined as in~(\ref{eq:FDR}). We term this reaction 'reversible'
since $\lambda_i$ can either increase or decrease. The motivation for the
postulated kinetics~\eqref{eq:chemical_reaction} is the physical
picture of a second-order reaction, where two molecules have to
interact for the dissociation process to occur. Of course, many
other kinetics are also possible, such as the first-order
irreversible evolution
$$
\frac{d\lambda_i}{dt} = K(\langle T_i \rangle)
(1-\langle \lambda_i \rangle),
$$
where~$\langle h \rangle = \sum_{j} \omega_r(r_{ij}) h_i$ denotes
a local spatial average. The choice of the reaction kinetics is
really a modelling choice, depending on the physical context.

The reaction constants $K_1$, $K_2$ are assumed to
depend only on internal temperatures of particles. For example, a
possible form in the Arrhénius spirit is:
\begin{equation}
\label{eq:cte_reac}
K_1(T) = Z_1 \ {\rm e}^{-E_1/k_{\rm B}T}, \qquad K_2(T) = Z_2 \ {\rm e}^{-E_2/k_{\rm B}T}.
\end{equation}
In this expression, $E_1$ and $E_2$ are the activation energies
of, respectively, the forward and backward reactions. In view of
our choice~\eqref{eq:chemical_reaction} of reaction kinetics, the
total increment of the progress variable is therefore the sum of
all elementary pair increments, which is very much in the DPD
spirit. The parameters~$Z_1,Z_2,E_1,E_2$ in the above equation can
be obtained from all-atom decomposition simulations or directly
from experiments. In particular, $Z_1$ and $E_1$ can be extracted
in the Arrhénius framework from the analysis of decomposition
rates at different initial temperatures. The constants 
$Z_2$ and $E_2$ can be
evaluated using thermochemical data (formation enthalpies).

For very exothermic reactions, $E_2 \gg E_1$, and
both energies are large since the activation energy is usually
large for energetic materials. The increment of a given reaction
rate is therefore non-negligible only if the material is sufficiently
heated. In practice, this can be achieved when a
strong shock is initiated in the system. If this shock is not
strong enough, chemical reactions do not occur fast enough, and
since the energy release is not sufficient, the shock wave is
weakened until it transforms into a sonic wave. On the contrary,
if the shock wave is strong enough, chemical reactions happen
close enough to the detonation front, and the energy released
sustains the shock wave such that it transforms into a
stationary detonation wave.
The progress of the reaction also
modifies the mechanical properties of the material. In particular,
reaction products usually have a larger specific volume than
reactants (at fixed thermodynamic conditions). Therefore, some
expansion is expected. The changes in the nature of the molecules
are taken into account by introducing two additional parameters
$k_a, k_E$ and using some mixing rule such as Berthelot's
rule. When the interaction potential is of Lennard-Jones form, the
interaction between the mesoparticles $i$ and $j$ is then given by
\begin{equation}
\label{eq:berthelot}
V(r_{ij}, \lambda_i, \lambda_j) = 4 E_{ij} \, \left ( \left (
\frac{a_{ij}}{r_{ij}} \right )^{12}  - \left ( \frac{a_{ij}}{r_{ij}}
\right)^6 \right ),
\end{equation}
with
$$
E_{ij} = E \sqrt{(1+k_E \lambda_i)(1+k_E \lambda_j)},
$$
and
$$
a_{ij} = a \left ( 1 + k_a
\frac{\lambda_i+\lambda_j}{2} \right ).
$$
When the reaction is
complete, the material initially described by a Lennard-Jones
potential of parameters $a, E$ is then described by a
Lennard-Jones of parameters $a' = a (1+k_a)$ and
$E' = E (1+k_E)$.


\section{Treatment of the exothermicity}
\label{sec:exothermicite}

We denote by $\Delta E_{\rm exthm}$ the exothermicity of the
reaction~(\ref{eq:reaction_chimique}). It is expected that $\Delta
E_{\rm exthm} = E_2 - E_1$. We assume that the
energy is liberated progressively during the reaction, in a manner
that the total energy of the system (chemical, mechanical,
internal) is preserved:
\begin{eqnarray*}
dH_{\rm tot}
& = &
d \left [  \sum_{i=1}^N \frac{p_i^2}{2m_i}
+ \epsilon_i + (1-\lambda_i) \Delta E_{\rm exthm}
+ \sum_{1 \leq i < j \leq N} V(r_{ij},\lambda_i,\lambda_j) \right ] = 0.
\end{eqnarray*}
In order to propose a dynamics satisfying this condition, we have
to make an additional assumption about the evolution of the
system. Neglecting diffusive processes, we require that, during
the elementary step corresponding to exothermicity, the total
energy of a given mesoparticle does not change\footnote{Of course,
during the elementary step corresponding to the
dynamics~(\ref{DPDsimplifie}), the total energy changes.}:
\begin{equation}
\label{eq:variations_total_energy}
d \left [ \frac12 \sum_{i \not = j} V(r_{ij},\lambda_i,\lambda_j)
  \right ] + d\left ( \frac{p_i^2}{2m_i} + \epsilon_i - \Delta E_{\rm
  exthm} \lambda_i\right ) = 0.
\end{equation}
We then consider evolutions of momenta and internal energies
balancing the variations in the total energy due to the variations
of $\lambda$ (exothermicity, changes in potential energies). This
is analogous to the fact that the variations of kinetic energy
in~(\ref{DPDsimplifie}) are compensated by variations of internal
energies. It is expected that the chemical energy liberated by the
reaction is converted into internal energy and kinetic energy. It
therefore remains to precise the quantitative distribution of the
energy. This is done using a predetermined ratio $0 < c < 1$.
This ratio could be measured in gas phase decomposition
experiments, using spectroscopy measurements (possibly numerical
simulations as well, by computing the temperature of internal
degrees of freedom after the reaction). 
Alternatively, it is possible
to postulate that~$c$ should be comparable to the ratio of the number
of external degrees of freedom for the products of the dissociation of
one complex molecule
(denoted by~$N_{\rm e}$),
divided by the total number of degrees of freedom~$N_{\rm t}$ in a
complex molecule (assuming some
equipartition of the liberated chemical energy). 
For exothermic chemical decompositions, it is expected that
a single complex molecule breaks into several smaller molecules
(still aggregated in a single mesoparticle), so that the above
ratio~$N_{\rm e}/N_{\rm t}$
should be close to 0.5.
Let us however emphasize at this point that numerical
profiles obtained here are robust enough with respect to the
choice of~$c$.

For the internal energies, the exothermicity of the reaction is
modelled as $d\epsilon_i = d{\cal E}_i$ with
$$
d{\cal E}_i = -c \left ( d \left [ \frac12 \sum_{i \not = j}
  V(r_{ij},\lambda_i,\lambda_j) \right ] - \Delta E_{\rm exthm}
d\lambda_i \right).
$$
For the momenta, we consider process ${\cal P}_i$ with
$dp_i = d{\cal P}_i$ such that
\begin{eqnarray*}
d\left( \frac{p^2_i}{2m} \right ) & = &
-(1-c) \left ( d \left [ \frac12
\sum_{i \not = j} V(r_{ij},\lambda_i,\lambda_j) \right ]  - \Delta
E_{\rm exthm} d\lambda_i \right).
\end{eqnarray*}
We explain in the next section how this is done in practice (see
Eq.~(\ref{eq:update_p_exo})).

Let us emphasize at this point that there are many other possible
ways to treat the exothermicity. For instance, it would be
possible to consider instantaneous reactions (jump processes for
which $\lambda$ changes from 0 to 1) occuring at random times, the
probability of reaction depending on the progress variable. However,
it is not clear whether such a dynamics is reversible, since the
reverse reaction requires particles to have large kinetic and
internal energies. In comparison, the process described here is
progressive and therefore, much more reversible.

Finally, we propose the following dynamics to describe reactive shock waves:
\begin{eqnarray}
\label{DPDreac}
dq_i & = & \frac{p_i}{m_i} \, dt, \nonumber \\
dp_i & = & \sum_{j, \, j \not = i} -\nabla_{q_i}
V(r_{ij},\lambda_i,\lambda_j) \, dt - \gamma_{ij} \chi^2(r_{ij})
v_{ij} \, dt + \sigma \chi(r_{ij}) dW_{ij} + d{\cal
  P}_{i}, \\
d\epsilon_i & =& \frac12 \sum_{j, \, j \not = i} \chi^2(r_{ij})
\left ( \gamma_{ij} v_{ij}^2 - \frac{d\sigma^2}{2}
\left(\frac{1}{m_i}+\frac{1}{m_j}\right) \right ) \, dt - \sigma \, \chi(r_{ij}) v_{ij} \cdot
dW_{ij} + d{\cal E}_i,
\nonumber \\
d\lambda_i & = & \sum_{j \not = i} \omega_r(r_{ij}) \left [
K_1(T_{ij}) ( 1 - \lambda_i)(1-\lambda_j)  + K_2(T_{ij}) \lambda_i
\lambda_j \right ] \, dt, \nonumber \\
\nonumber
\end{eqnarray}
where $d{\cal P}_i$, $d{\cal E}_i$ are such
that~(\ref{eq:variations_total_energy}) holds, {\it i.e.} the
total energy is conserved. The fluctuation-dissipation relation
relating $\gamma_{ij}$ and $\sigma$ is still~\eqref{eq:FDR}.
Notice also that the inert
dynamics~(\ref{DPDsimplifie}) of~\cite{Stoltz06} is recovered when
$Z_1 = Z_2 = 0$, starting from $\lambda_i = 0$ for
all $i$.


\section{Numerical implementation}
\label{sec:implementation}

The numerical integration of~(\ref{DPDreac}) is done using a
decomposition of the dynamics into elementary stochastic
differential equations. We denote by $\phi^t_{\rm inert}$ the flow
associated with the dynamics~(\ref{DPDsimplifie}), and by
$\phi^t_{\rm reac}$ the flow associated with the remaining part of
the dynamics~(\ref{DPDreac}): $\forall 1 \leq i \leq N$,
\begin{equation}
\label{eq:generator_reaction}
\left \{ \begin{array}{cl}
d\lambda_i = & \sum_{j \not = i} \omega_r(r_{ij}) \left [ K_1(T_{ij})
  ( 1 - \lambda_i)(1-\lambda_j)  + K_2(T_{ij}) \lambda_i \lambda_j \right ] \, dt, \\
dp_i = & d{\cal P}_i, \\
d\epsilon_i = & d{\cal E}_i. \\
\end{array} \right.
\end{equation}
A one-step integrator for a time-step $\Delta t$ is constructed as
$$
(q^{n+1},p^{n+1},\epsilon^{n+1},\lambda^{n+1}) = \Phi^{\Delta t}_{\rm
  reac} \circ \Phi^{\Delta t}_{\rm
  inert}(q^{n},p^{n},\epsilon^{n},\lambda^{n}).
$$
A possible numerical flow $\Phi^{\Delta t}_{\rm inert}$ is given in~\cite{Stoltz06}.

Let us now construct a numerical flow $\Phi^{\Delta t}_{\rm reac}$
approximating the flow $\phi^{\Delta t}_{\rm reac}$. Denoting
$(q^{n+1},\tilde{p}^{n},\tilde{\epsilon}^{n},\lambda^{n})
=\Phi^{\Delta t}_{\rm inert}(q^n,p^n,\epsilon^n,\lambda^n)$, we
first integrate the evolution equation on the progress
variables~$\lambda_i$ using a first-order explicit integration:
\begin{eqnarray*}
\tilde{\lambda}_i^{n+1} & = & \lambda_i^{n} +  \sum_{j \not = i}
  \omega_r(r^{n+1}_{ij}) \left [ K_1(\tilde{T}^n_{ij}) ( 1 -
  \lambda^n_i)(1-\lambda^n_j) K_2(\tilde{T}^n_{ij})
\lambda^n_i \lambda^n_j \right ] \ \Delta t. \nonumber
\end{eqnarray*}
We then set $\lambda_i^{n+1} =
\min(\max(0,\tilde{\lambda}_i^{n+1}),1)$ in order to ensure that
the progress variable remains between~0 and~1. Once all progress variables
are updated, the variation~$\delta E_i^n$ in the total energy of
particle $i$ due to the variations of $\{\lambda_j\}$ is computed
as
\begin{eqnarray*}
\delta E_i^n & = & (\lambda_i^{n+1}-\lambda_i^n) \Delta E_{\rm exthm}
 + \frac12 \sum_{j \not = i} \left ( V(r^{n+1}_{ij},\lambda_i^{n+1},\lambda_j^{n+1}) - V(r^{n+1}_{ij},\lambda_i^{n},\lambda_j^{n}) \right ).
\end{eqnarray*}
The conservation of total energy is then ensured through
variations of internal and kinetic energies. The internal energies
are updated as $\epsilon_i^{n+1} = \tilde{\epsilon}_i^n + c \,
\delta E_i^n$. The update of $p_i^{n+1}$ is done by adding to
$p_i^n$ a vector with random direction, so that the final
momentum is such that the kinetic energy is correct. More
precisely, when the dimension of the physical space is $d=2$ for
example, an angle $\theta_i^n$ is chosen at random in the interval
$[0,2\pi]$, the angles $(\theta_i^n)_{i,n}$ being independent and
identically distributed random variables. The new
momentum $p_i^{n+1}$ is then constructed such that
\begin{eqnarray}
\label{eq:update_p_exo}
p_i^{n+1} = p_i^n + \alpha^n (\cos \theta_i^n, \sin \theta_i^n), \nonumber
\\
\frac{(p_i^{n+1})^2}{2m_i} = \frac{(\tilde{p}_i^{n})^2}{2m_i} + (1-c) \, \delta E_i^n.
\end{eqnarray}
Solving this equation in $\alpha^n$ gives the desired result.


\section{Numerical results}

We present in this section numerical results obtained for the
dynamics~(\ref{DPDreac}) of a two-dimensional fluid. A shock is
initiated using a piston of velocity~$u_{\rm p}$ during a
time~$t_{\rm p}$. The initial conditions for the positions~$q_i$,
momenta~$p_i$ and internal energies~$\epsilon_i$ are sampled as
proposed in~\cite{Stoltz06}.

\begin{figure*}
\begin{center}
\includegraphics[angle=270,width=15cm]{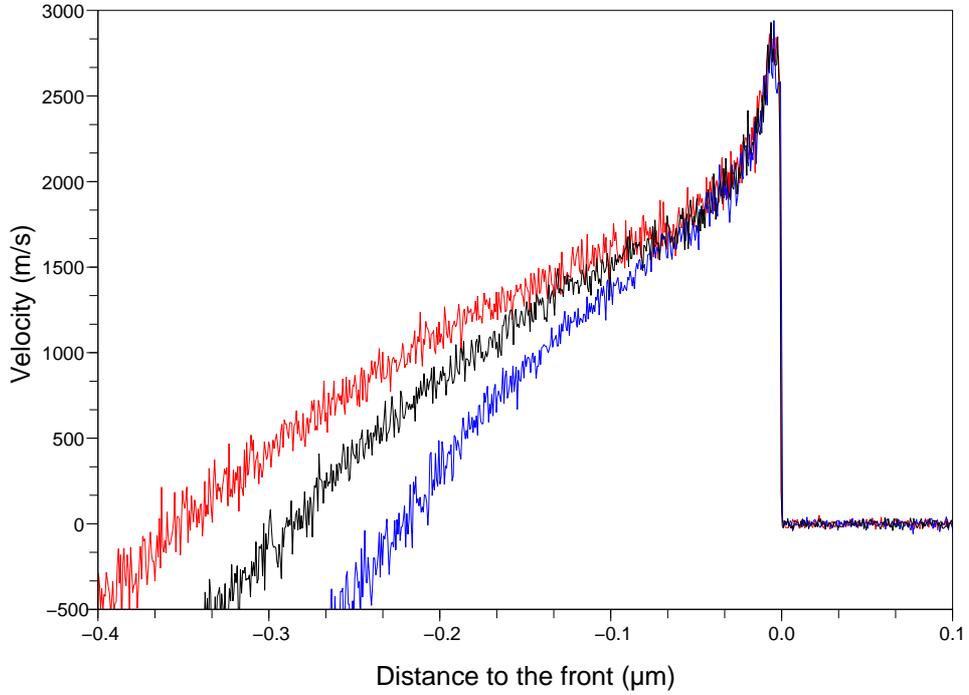}
\caption{\label{fig:autosimilarite} (color online) Velocity profiles
  in the material at different times, with shock fronts aligned (Black: $t=1.6 \times
  10^{-10}$~s; Blue: $t=2.0 \times 10^{-10}$~s; Red: $t=2.4 \times
  10^{-10}$~s).}
\end{center}
\end{figure*}

We consider the following parameters, inspired by the nitromethane
example (in which case the molecule~CH$_3$NO$_2$ would replaced by a mesoparticle in a space of 2 dimensions).
The parameters can be classified in four main categories, depending on
whether they describe the mechanical properties of the
material, characterize
the inert dynamics and the chemical kinetics,
or are related to the exothermicity. We
consider here a system with
\begin{itemize}
\item (Material parameters) a molar mass $m=80$~g/mol,
described by a
Lennard-Jones potential of parameter $E_{\rm LJ} = 3 \times
10^{-21}$~J (melting temperature around 220~K) and $a = 5$~\AA, with a
cut-off radius $r_{\rm cut} = 15$~\AA~for the
computation of forces. The changes in the parameters of the
Lennard-Jones material during the reaction
follow~(\ref{eq:berthelot}), using $k_E = 0$ and $k_a = 0.2$
(pure expansion).
\item (Parameters of the inert dynamics)
The microscopic state law is $\epsilon = C_v T$
with $C_v = 10$~$k_{\rm B}$ ({\it i.e.}, 20 degrees of freedom are not
represented). The friction is $\gamma =
10^{-15}$~kg/s, and the dissipation weighting function $\chi(r) =
(1-r/r_{\rm c})$, with $r_{\rm c} = r_{\rm cut}$.
\item (Chemical kinetics)
For the chemical reaction~(\ref{eq:chemical_reaction}), reaction
constants are computed using~(\ref{eq:cte_reac}) with $Z_1 =
Z_2 = 10^{17}$~s$^{-1}$, $E_1/k_{\rm B} = 15000$~K, the
exothermicity being $\Delta E_{\rm exthm} = 6.25$~eV. The reaction
weighting function $\omega(r) = \chi(r)$;
\item (Exothermicity) we choose $c=0.5$.
\end{itemize}
The initial density of the system is $\rho = 1.06$~g/cm$^{-3}$,
and the initial temperature $\bar{T}=300$~K. The time-step used is
$\Delta t = 2 \times 10^{-15}$~s. Figure~\ref{fig:autosimilarite}
presents velocity profiles averaged in thin slices of the material
in the direction of the shock, for a compression time $t_{\rm p} =
2$~ps at a velocity $u_{\rm p} = 5000$~m/s. We tested the
independence of velocity profiles in the reaction zone with
respect to different initial loadings $(t_{\rm p}, u_{\rm
p})=(1~{\rm ps}, 6000~{\rm m/s})$, $(t_{\rm p}, u_{\rm p})=(2~{\rm
ps}, 6000~{\rm m/s})$, $(t_{\rm p}, u_{\rm p})=(3~{\rm ps},
6000~{\rm m/s})$ and~$(t_{\rm p}, u_{\rm p})=(3~{\rm ps},
5000~{\rm m/s})$.

\begin{figure*}
\includegraphics[angle=270,width=7.5cm]{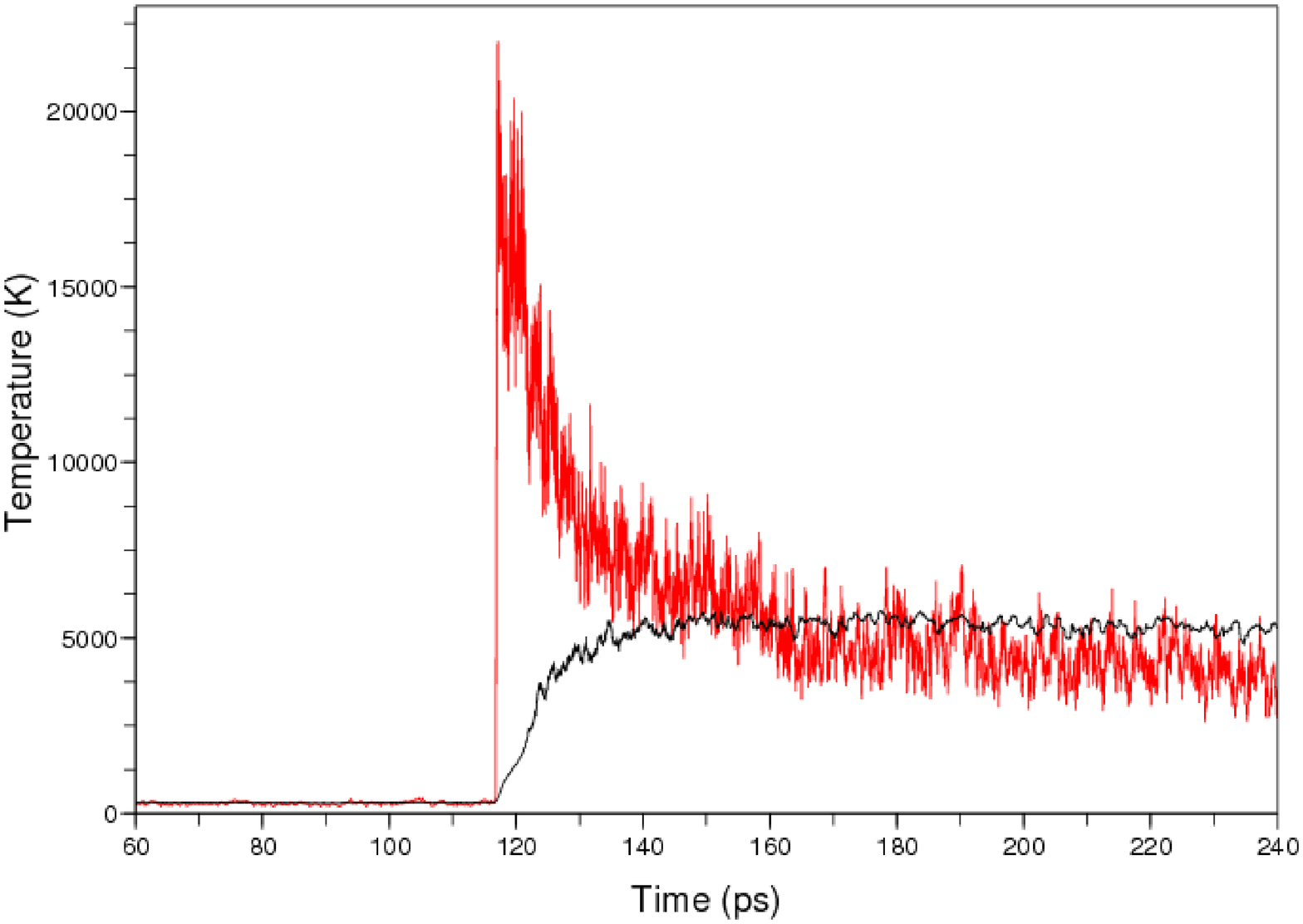}\hfill
\includegraphics[angle=270,width=7.5cm]{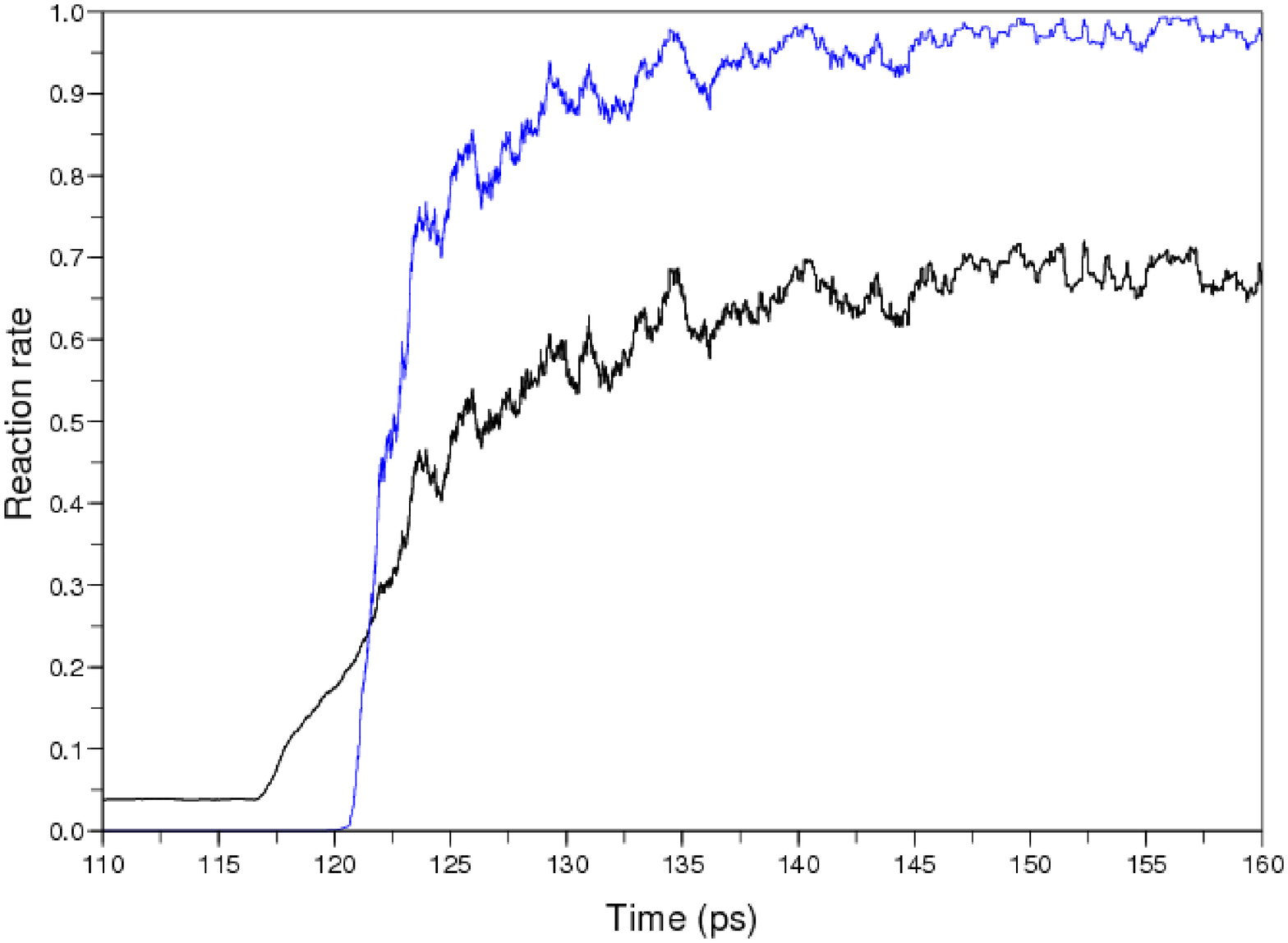}
\caption{\label{fig:slice_autosim} (color online) Left : variations of internal
(red) and kinetic (black) temperatures in the direction of the
shock, as a function of time in a slice of material. Right:
evolution of the progress variable averaged in a slice of material as
a function of time (blue). For comparison, a rescaled internal
temperature profile is also presented (black).}
\end{figure*}

The velocity of the shock front is constant, and approximately
equal to $u_{\rm s} = 1820$~m/s. Notice that the wave can be
divided into three regions: the upstream region is unperturbed;
the region around the shock front where chemical reactions happen
is of constant width (approximately~300-400~\AA, which is
consistent with all-atoms studies, see for
instance~\cite{HGGKHS06}); the downstream region is an autosimilar
rarefaction wave. This profile is therefore reminiscent of ZND
profiles~\cite{FD} encountered in hydrodynamic simulations of
detonation waves.

Figure~\ref{fig:slice_autosim} presents the evolution of internal
and kinetic temperatures averaged in a slice of material in the
direction of the shock as a function of time (Left), as well as
the evolution of the average progress variables (Right). In
particular, the reaction does not start immedialely at the shock
front: the ignition asks first for a sufficient heating of the
material (through an increasing internal energy), since the
reaction constant are too low at temperatures lower than a few
thousands Kelvins with the values chosen here.


\section{Conclusion and perspectives}

In conclusion, we extended the mesoscopic inert model for shock
waves of~\cite{Stoltz06} to the reactive case. The key idea is to
introduce an additional variable describing the progress of the
reaction, and to postulate an activated exothermic chemical
reaction. Provided the initial loading is strong enough, the
energy released by chemical reactions sustains the shock wave,
whose front then travels unperturbed in the material, followed by
a rarefaction wave.

Now that while the qualitative behavior of the model is granted, more
quantitative studies must be pursued, where a careful comparison
between all-atom numerical results and the profiles predicted by
the reduced model proposed here are to be compared. Such studies
were already performed in the inert case~\cite{SH04,Stoltz06}.

A promising track for a real multiscale modelling of detonation
waves would now be to further reduce the model proposed here, in
order to obtain a description of matter at a scale truly
intermediate between SPH and all-atom models. In such a model,
random fluctuations would still be present, though of much lower
magnitude; on the other hand, local thermodynamic fields (such as
pressure) should also be introduced. Such an intermediate model is
necessary to describe mesoscopic effects having an impact on the
macroscopic features of the detonation, such as hot spot ignition.

%
%

\newcommand{\Name}[1]{{\scshape #1},}
\newcommand{\Review}[1]{{\itshape #1},}
\newcommand{\Vol}[1]{{\bfseries #1}}
\newcommand{\Page}[1]{#1}
\newcommand{\Year}[1]{(#1)}
\newcommand{\Book}[1]{{\itshape #1}}
\newcommand{\Publ}[1]{{\normalfont(#1)}}
\newcommand{\REVIEW}[4]{\Review{#1} \Vol{#2} \Year{#3} \Page{#4}}
\newcommand{\SAME}[3]{\Vol{#1} \Year{#2} \Page{#3}}

\end{document}